# Kinetic and Electrostatic Energies in Quantum Mechanics

Y Kornyushin

Maître Jean Brunschvig Research Unit, Chalet Shalva, Randogne, CH-3975

E-mail: jacqie@bluewin.ch

**Abstract.** A concept of kinetic energy in quantum mechanics is analyzed. Kinetic energy is a non-zero positive value in many cases of bound states, when a wave function is a real-valued one and there are no visible motion and flux. This can be understood, using expansion of the wave function into Fourier integral, that is, on the basis of virtual plane waves. The ground state energy of a hydrogen atom is calculated in a special way, regarding explicitly all the terms of electrostatic and kinetic energies. The correct values of the ground state energy and the radius of decay are achieved only when the electrostatic energies of the electron and the proton (self-energies) are not taken into account. This proves again that self-action should be excluded in quantum mechanics. A model of a spherical ball with uniformly distributed charge of particles is considered. It is shown that for a neutral ball (with compensated electric charge) the electrostatic energy is a non-zero negative value in this model. This occurs because the self-energy of the constituting particles should be subtracted. So it shown that the energy of the electric field does not have to be of a positive value in any imaginable problem.

## 1. Introduction

Let us write down Schrödinger equation for an elementary particle [1]:

$$-(\hbar^2/2m)\Delta\psi + U(\mathbf{r})\psi = E\psi, \tag{1}$$

where $\hbar$ is Planck constant, divided by $2\pi$, $m$ is the mass of a particle, $\Delta$ is Laplace operator, $\psi$ is a wave function, $U(\mathbf{r})$ is the potential energy of a particle as a function of a radius-vector $\mathbf{r}$, and $E$ is the energy of the stationary state of a particle. In a case of a free particle $U(\mathbf{r}) \equiv 0$ and Eq. (1) has a well-known solution, a plane wave [1], $\psi(\mathbf{r}) = \text{const}[\exp i(\mathbf{k},\mathbf{r})]$, where $i$ is the imaginary unity, and $\mathbf{k}$ is the wave-vector. This solution corresponds to a non-localized free particle. The energy of this particle is well known to be $E(k) = \hbar^2 k^2/2m$ [1]. This is a kinetic energy of a non-localized particle [1].

Potential energy $U(\mathbf{r})$ in Schrödinger equation, Eq. (1), does not include the potential energy of the interaction of the considered elementary particle with the field, produced by this same particle. Thus, in quantum mechanics the action of the elementary particle on itself, that is a self-action, is not taken into account. Averaging Eq. (1) quantum-mechanically we get: $E = \langle T \rangle + \langle U(\mathbf{r}) \rangle$, where $T = -(\hbar^2/2m)\Delta$ is the kinetic energy operator. We see that the energy of the stationary state of a particle does not include the potential energy of the particle itself. Let us consider a charged particle, which produces electric field. The energy of this field is not included in the energy of the stationary state of a particle. This is because the energy of the electric field, produced by the particle, is not present in Schrödinger equation. It should be noted that this refers only to the particles charged with elementary charge $\pm e$, such as electron, proton, positron, antiproton. When regarded particle consists of, for example, two protons, the field, produced by each proton, is an external one with respect to the other proton, and the potential energy of the interaction of the two protons should be introduced in Schrödinger equation. Thus, the charged elementary particle does not posses electrostatic energy in quantum mechanics.

When a charged particle is localized, it has a kinetic energy and creates an electric field. The energy stored in this field should not be taken into account when the total energy of a system is calculated, as it would be a self-action. So here we have electric field with no stored energy.

We shall discuss the kinetic energy when there is no flux/current, the electrostatic field without any electrostatic energy stored in it, and electrostatic energy when there is no electrostatic field. It shows again and again how wonderful quantum mechanics is.

**2. Kinetic energy in a general case**

Kinetic energy is the energy of a motion [1]. A matter flux and electric current accompany the motion of a charged matter. Matter flux **J** is given by the following equation [1]:

$$\mathbf{J} = (i\hbar/2m)(\psi\nabla\psi^* - \psi^*\nabla\psi). \qquad (2)$$

It seems that no motion without a flux of a matter exists. In many of the bound states (both ground states and excited ones) of any particle, e.g., electron, the wave functions are real-valued, they contain no imaginary parts [1]. As follows from Eq. (2), the real-valued wave function yields no flux and no electric current for a charged particle. As an example let us take an electron in the ground state of a hydrogen atom. Corresponding wave function is [1] $\psi(r) = (g^3/\pi)^{1/2}\exp{-gr}$, where $g = me^2/\hbar^2$ and $r$ is the distance from the proton [1].

The operator of the momentum is $-i\hbar\nabla$, where $\nabla$ is the gradient operator. The kinetic energy operator $T$ is the operator of the momentum in square, divided by $2m$ [1]. This operator, $-(\hbar^2/2m)\Delta$, averaged quantum-mechanically over the whole space, is an average kinetic energy, $\langle T \rangle$ [1]. From this follows that for the electron in the ground state of a hydrogen atom $\langle T \rangle = \langle T \rangle_0 = me^4/2\hbar^2$. So we have here a very essential kinetic energy with no visible motion. This can be explained using virtual plane waves expansion.

As was mentioned above, the operator $-(\hbar^2/2m)\Delta$, averaged over the whole space yields kinetic energy. This energy is not zero because the wave function is inhomogeneous. It depends on coordinates. Let us expand the real-valued wave function into Fourier integral:

$$\psi(r) = \int a_{\mathbf{k}}[\exp i(\mathbf{k},\mathbf{r})]d\mathbf{k}. \qquad (3)$$

The expansion coefficients are normalized, so that

$$\int a^*_{\mathbf{k}} a_{\mathbf{k}} d\mathbf{k} = 1. \qquad (4)$$

The kinetic energy operator $-(\hbar^2/2m)\Delta$, averaged over the whole space, is the kinetic energy of a system. Using Eq. (3) we have:

$$\langle T \rangle = (\hbar^2/2m)\int k^2 a^*_\mathbf{k} a_\mathbf{k} d\mathbf{k}. \tag{5}$$

Equation (5) shows that the kinetic energy is a sum of kinetic energies of constituting plane waves with the appropriate weight.

**3. Hydrogen atom**

Let us consider in this section a simple problem of a free hydrogen atom in a ground state. Let us do it in an alternative way, like it was done in [2]. Let us write the wave function of the electron in the ground state of a hydrogen atom in a form [2]

$$\psi(r) = [\exp(-r/2R)]/2(2\pi)^{1/2} R^{3/2}, \tag{6}$$

where $R$ is related to the equilibrium (or not) radius of decay.

The energy of the ground state of the electron in a hydrogen atom is described by a well-known relation [1]:

$$E_{0h} = -(me^4/2\hbar^2), \tag{7}$$

where $e$ is the electron (negative) charge and $m$ is the electron mass.

The electrostatic field created by the electron shell with the electric charge density

$$\rho(r) = e\psi^*(r)\psi(r) = (e/8\pi R^3)\exp(-r/R), \tag{8}$$

can be calculated using Gauss theorem. The corresponding electrostatic potential is

$$\varphi(r) = (e/r) - (e/2R)[(2R/r) + 1]\exp(-r/R). \tag{9}$$

In the vicinity of $r = 0$ the electrostatic potential is as follows:

$$\varphi(r) \approx (e/2R) - (e/12R^3)r^2 + \ldots . \tag{10}$$

Kinetic energy of the electron is a quantum-mechanically averaged value of its operator:

$$T = -(\hbar^2/2m)\Delta. \tag{11}$$

It follows from Eqs. (6) and (11) that

$$\langle T \rangle = \hbar^2/8mR^2. \tag{12}$$

In quantum mechanics there is no self-action. Hence, the electrostatic energy of an elementary particle does not exist in quantum mechanics. In the simple problem of a hydrogen atom in a ground state there is no electrostatic energy of the electron and the proton. The only electrostatic energy that exists really is the energy of electron-proton interaction. This energy is negative, because of the opposite charges of the interacting particles. So, hydrogen atom is a very interesting object, which has electrostatic field and negative electrostatic energy. Such a situation is never possible in classical electrostatics. In classical electrostatics the energy stored in a total electric field is proportional to the integral of this total electric field in square over the whole space, and it is always positive.

The total energy of a hydrogen atom (without self-action), which consists of the kinetic energy of the electron, Eq. (12), and the electrostatic interaction energy of the positively charged core with the potential in the center of the electronic shell, Eq. (10), is as follows:

$$E_h(R) = (\hbar^2/8mR^2) - (e^2/2R). \tag{13}$$

As was mentioned above, the electrostatic energy of the electron itself should not be taken into account as it represents the so-called self-action. The same should be said about the electrostatic energy of the proton. Energy, described by Eq. (13) has a minimum at $R = R_e = \hbar^2/2me^2$, as is well known [1]. Correct minimum value of the ground state energy is given by Eq. (7). Corresponding value of the electrostatic energy is $U_0 = -(e^2/2R_e) = -(me^4/\hbar^2)$. This value is the correct one. Indeed, as $E_{0h} = U_0 + \langle T \rangle_0$, $U_0 = E_{0h} - \langle T \rangle_0$, $\langle T \rangle_0 = (me^4/2\hbar^2)$ (as was mentioned above), and $E_{0h} = -(me^4/2\hbar^2)$ (see Eq. (7)), we have $U_0 = -(me^4/\hbar^2)$.

Here we calculated the energy of the ground state of a hydrogen atom in a special way, regarding all the terms of the electrostatic and kinetic energy. We did it on purpose. Thus we see explicitly, that the energy of the ground state of a hydrogen atom, calculated here without taking into account the electrostatic energies of the electron and the proton (that is, without self-action), has a correct well known and measured experimentally value [1].

A problem of negative hydrogen ion was considered in [3].

**4. Positronium**

Now let us imagine a positronium (a hydrogen type atom consisting of electron and positron) in a ground state. The masses of the both particles are of the same value. Their electric charges are of the same value, but of the opposite signs. Because of this the wave functions of both particles in a ground state are identical. So here we have no local charge, no electrostatic field but we have negative electrostatic energy of a system.

**5. Uniformly charged spherical ball model**

Let us consider in this section $N$ electrons with the electric charge of each one distributed uniformly in a spherical ball of a radius $R$. Electrostatic energy of such a ball in classical electrostatics is [4] $U = 0.6e^2N^2/R$. Electrostatic energy of each of the electron is $U = 0.6e^2/R$. Electrostatic energy of $N$ electrons is $U = 0.6e^2N/R$. In quantum mechanics this energy should be subtracted from the classical energy of a system. So, the electrostatic energy of such a ball in quantum mechanics is [4]:

$$U = 0.6e^2N(N-1)/R. \tag{14}$$

Self-action exists in classical electrostatics. So we have $N^2$ instead of $N(N-1)$ in Eq. (14) in classical electrostatics.

For a spherical ball of a radius *R*, containing *N* protons with uniformly distributed over the ball electric charge of each one we have the same equation, Eq. (14), for the electrostatic energy.

Let us consider now a spherical ball of a radius *R*, containing *N* electrons and *N* protons, each one of them distributed uniformly inside the ball. The total charge in any point of a ball is zero. So there is no electrostatic field. The total electrostatic energy of such a ball consists of the electrostatic energy of electrons, Eq. (14), electrostatic energy of protons, Eq. (14), and interaction energy, $U_i = -1.2e^2N^2/R$. So the total electrostatic energy is described by the following expression:

$$U_t = -1.2e^2N/R. \qquad (15)$$

This result is a particular case of more general expressions, published in [4].

In classical electrostatics, as was mentioned, we have $N^2$ instead of $N(N-1)$ in Eq. (14). Hence we have the total electrostatic energy zero. In quantum mechanics, where self-action does not exist, we have negative total electrostatic energy in a locally electrically neutral system. This occurs because every particle produces electrostatic field and in spite of the local neutrality we should subtract from zero the total electrostatic energy of each separate particle. Thus we come to the negative value of the electrostatic energy.

In this section we have considered a case of no local electric charge, no local electrostatic field and a considerable negative electrostatic energy.

## 6. Discussion

The concept of the kinetic energy in non-relativistic quantum mechanics is considered here. Kinetic energy is a non-zero positive value for many cases with no visible motion and flux. The momentum operator contains a factor of imaginary unity *i*. So when a wave function is a real-valued one, the quantum-mechanically averaged momentum, **p**, is imaginary, $\mathbf{p} = -i\hbar\int\psi\nabla\psi dv$. But using the expansion of the real-valued wave function into Fourier integral, Eq. (3), we have $\mathbf{p} = \hbar\int\mathbf{k}a^*_\mathbf{k}a_\mathbf{k}d\mathbf{k}$, which is a real-valued quantity. From this follows that for the real-valued wave function $\mathbf{p} = 0$, as it is the only possibility to be real and imaginary at the same time.

The existence of a positive kinetic energy in the states with no flux can be understood, using expansion in virtual plane waves. This expansion presents the kinetic energy as a sum of partial kinetic energies of constituting virtual plane waves. The real-valued wave function is a superposition of moving plane-waves. Their velocities cancel, but their kinetic energies add.

The energy of the ground state of a hydrogen atom is calculated in a special way, considering all the terms of the electrostatic and kinetic energy of the electron and the proton. It is shown that the correct values of the energy of the ground state and the radius of decay could be obtained only when the electrostatic energy of the electron and the proton are not taken into account. This supports once more the concept that self-action should not be taken into account in quantum mechanics.

When self-action in quantum mechanics is not taken into account we have some cases of zero local charge, but non-zero negative electrostatic energy. This occurs, in particular, in a positronium and in the model of a spherical ball, consisting of an equal number of positive and negative elementary particles. This model is also considered in this paper.

## 7. Conclusions

Simple methodical approach of expanding wave function into a series of virtual plane waves occurred to be helpful for understanding the nature of the kinetic energy of a particle, described by real-valued wave function.

It was shown also that the energy of the electric field does not have to be of a positive value in any imaginable problem.

## Acknowledgements

I would like to thank Kirk McDonald of Princeton University, Norman H. Redington of MIT, Mark Dykman of Michigan State University, and Miron Amusia of A. F. Ioffe Institute in Sankt-Petersburg for useful discussions.

## References

[1] Landau L D and Lifshitz E M 1987 *Quantum Mechanics* (Oxford: Pergamon)
[2] Kornyushin Y 2003 *Facta Universitatis Series PCT* **2** 253
[3] Amusia M Y and Kornyushin Y 2000 *Eur. J. Phys.* **21** 369
[4] Amusia M Y and Kornyushin Y 2000 *Contemporary Physics* **41** 219

# Кинетическая и электростатическая энергии в квантовой механике

### Ю. В. Корнюшин

Maître Jean Brunschvig Research Unit, Chalet Shalva, Randogne, CH-3975

mailto: jacqie@bluewin.ch

**Резюме.** Анализируется понятие кинетической энергии в квантовой механике. Кинетическая энергия является положительной, не равной нулю величиной во многих связанных состояниях, когда волновая функция вещественна, отсутствует поток и как-будто нет движения. Это можно понять, если представить волновую функцию в виде интеграла Фурье, то есть, на основе виртуальных плоских волн. Энергия основного состояния атома водорода вычислена с явным учетом всех слагаемых кинетической и электростатической энергий. Правильные величины энергии основного состояния и его радиуса получаются только в том случае, когда собственные электростатические энергии электрона и протона (самодействия) исключаются. Это ещё раз доказывает, что самодействие в квантовой механике отсутствует. Рассмотрена модель шара с однородно распределенным зарядом частиц. Показано, что в этой модели электростатическая энергия нейтрального (с компенсированным зарядом) шара является отличной от нуля отрицательной величиной. Таким образом показано, что энергия электрического поля не является во всех случаях обязательно положительной величиной.

## 1. Введение

Рассмотрим уравнение Шредингера [1]:

$$-(\hbar^2/2m)\Delta\psi + U(\mathbf{r})\psi = E\psi, \qquad (1)$$

где $\hbar$ – постоянная Планка, деленная на $2\pi$, $m$ – масса частицы, $\Delta$ – оператор Лапласа, $\psi$ – волновая функция, $U(\mathbf{r})$ – потенциальная энергия частицы, $\mathbf{r}$ – радиус-вектор и $E$ – энергия стационарного состояния частицы. В случае свободной частицы $U(\mathbf{r}) \equiv 0$ и уравнение (1) имеет общеизвестное решение, плоскую волну [1], $\psi(\mathbf{r}) = \text{const}[\exp i(\mathbf{k},\mathbf{r})]$, где $i$ – мнимая единица и $\mathbf{k}$ – волновой вектор. Это решение соответствует нелокализованной свободной частице. Энергия этой частицы хорошо известна, $E(k) = \hbar^2 k^2/2m$ [1]. Это есть кинетическая энергия свободной частицы.

Потенциальная энергия $U(\mathbf{r})$ в уравнении Шредингера (1) не содержит потенциальную энергию взаимодействия рассматриваемой элементарной частицы с полем, которое она сама производит. Таким образом, в квантовой механике действие элементарной частицы на саму себя, то есть самодействие, не должно учитываться. Усредняя квантово-механически уравнение (1), имеем: $E = \langle T \rangle + \langle U(\mathbf{r}) \rangle$, где $T = -(\hbar^2/2m)\Delta$ есть оператор кинетической энергии. Видим, что энергия стационарного состояния частицы не содержит потенциальную энергию самой частицы. Рассмотрим заряжённую частицу, создающую электрическое поле. Энергия этого поля не содержится в энергии стационарного состояния частицы, так как энергия поля, создаваемого частицей не присутствует в уравнении Шредингера. Следует отметить что это относится только к частицам, заряжённым элементарным зарядом $\pm e$, таким как электрон, протон, позитрон, антипротон. Когда рассматриваемая частица состоит, например, из двух протонов, поле, производимое каждым протоном, является внешним по отношению к другому протону и потенциальная энергия взаимодействия двух протонов должна присутствовать в уравнении Шредингера. Отсюда вытекает, что элементарная заряженная частица не обладает потенциальной электростатической энергией в квантовой механике.

Локализованная заряженная частица создает электрическое поле и обладает кинетической энергией. При вычислении полной электростатической энергии системы энергия электростатического поля частицы не должна учитываться, так как это было бы учетом самодействия. Это значит, что здесь мы имеем электростатическое поле не обладающее энергией.

Мы обсудим случаи положительной кинетической энергии при отсутствии потока (тока), электростатическое поле не обладающее энергией, а также электростатическую энергию в отсутствие электростатического поля. Это показывает ещё раз насколько квантовая механика удивительна.

**2. Кинетическая энергия в общем случае**

Кинетическая энергия – это энергия движения [1]. Движение заряженной материи сопровождается потоком вещества и электрическим током. Поток вещества $\mathbf{J}$ описывается следующим уравнением [1]:

$$\mathbf{J} = (i\hbar/2m)(\psi\nabla\psi^* - \psi^*\nabla\psi), \qquad (2)$$

где $\nabla$ – оператор градиента.

Кажется, что без потока вещества нет движения. Во многих связанных состояниях (как основных, так и возбуждённых) частицы (например, электрона) волновые функции вещественны, не содержат мнимой части [1]. Как следует из (2), вещественная волновая функция заряженной частицы не дает потока вещества и электрического тока. Например, волновая функция основного состояния электрона в атоме водорода [1] $\psi(r) = (g^3/\pi)^{1/2}\exp{-gr}$, где $g = me^2/\hbar^2$ (здесь $e$ — отрицательный элементарный заряд электрона и $r$ — расстояние от протона). Эта волновая функция вещественна.

Оператор импульса есть $-i\hbar\nabla$. Оператор кинетической энергии $T$ есть оператор импульса в квадрате, деленный на $2m$ [1]. Этот оператор, $-(\hbar^2/2m)\Delta$, усредненный квантово-механически по всему пространству, есть средняя кинетическая энергия, $\langle T \rangle$ [1]. Так, средняя кинетическая энергия электрона в основном состоянии атома водорода $\langle T \rangle = \langle T \rangle_0 = me^4/2\hbar^2$. То есть в этом случае имеется существенная кинетическая энергия без видимого движения. Это можно понять с помощью разложения в ряд по виртуальным плоским волнам.

Как было отмечено выше, оператор $-(\hbar^2/2m)\Delta$, усредненный квантово-механически по всему пространству есть кинетическая энергия. Эта энергия отлична от нуля потому что волновая функция неоднородна, она зависит от координат. Разложим вещественную волновую функцию в интеграл Фурье:

$$\psi(r) = \int a_{\mathbf{k}}[\exp i(\mathbf{k},\mathbf{r})]d\mathbf{k}. \qquad (3)$$

Коэффициенты разложения нормализованы так, что

$$\int a^*_{\mathbf{k}} a_{\mathbf{k}} d\mathbf{k} = 1. \qquad (4)$$

Оператор кинетической энергии $-(\hbar^2/2m)\Delta$, усредненный по всему пространству, есть кинетическая энергия. С помощью (3) получаем:

$$\langle T \rangle = (\hbar^2/2m)\int k^2 a^*_{\mathbf{k}} a_{\mathbf{k}} d\mathbf{k}. \qquad (5)$$

Уравнение (5) показывает, что кинетическая энергия является суммой кинетических энергий составляющих плоских волн с соответствующим каждой плоской волне весом.

## 3. Атом водорода

В этом разделе рассмотрим простую задачу об основном состоянии электрона в атоме водорода. Рассмотрим эту задачу специальным образом, как это было сделано в [2]. Запишем волновую функцию основного состояния в виде [2]

$$\psi(r) = [\exp(-r/2R)]/2(2\pi)^{1/2}R^{3/2}, \qquad (6)$$

где $R$ обозначает равновесный или неравновесный радиус волновой функции.

Энергия основного состояния электрона в атоме водорода [1]

$$E_{0h} = -(me^4/2\hbar^2). \qquad (7)$$

Электростатическое поле, создаваемое зарядом электрона с плотностью

$$\rho(r) = e\psi^*(r)\psi(r) = (e/8\pi R^3)\exp(-r/R), \qquad (8)$$

можно вычислить с помощью теоремы Остроградского-Гаусса.

Соответствующий электростатический потенциал

$$\varphi(r) = (e/r) - (e/2R)[(2R/r) + 1]\exp(-r/R). \qquad (9)$$

В окрестности $r = 0$ электростатический потенциал имеет вид:

$$\varphi(r) \approx (e/2R) - (e/12R^3)r^2 + \ldots . \qquad (10)$$

Кинетическая энергия электрона является квантово-механическим средним оператора:

$$T = -(\hbar^2/2m)\Delta. \qquad (11)$$

Из уравнений (6) и (11) следует:

$$\langle T \rangle = \hbar^2/8mR^2. \qquad (12)$$

В квантовой механике самодействие отсутствует. Поэтому электростатическая энергия элементарных частиц также равна нулю. В простой задаче об основном состоянии атома водорода электростатические энергии электрона и протона также равны нулю. Остается только электростатическая энергия взаимодействия электрона и протона. Эта энергия отрицательна, так как знаки зарядов электрона и протона противоположны. Таким образом, атом водорода является очень интересным объектом с электростатическим полем и отрицательной электростатической энергией. В классической электростатике такое совершенно невозможно, так как электростатическая энергия пропорциональна интегралу по всему пространству от квадрата полного электростатического поля и результат всегда положительный.

Полная энергия атома водорода (без самодействия) состоит из кинетической энергии электрона и электростатической энергии взаимодействия положительно заряженного ядра с отрицательно заряженным электронным облаком. Эта энергия, как следует из (10) и (12), описывается следующим выражением:

$$E_h(R) = (\hbar^2/8mR^2) - (e^2/2R). \qquad (13)$$

Как было отмечено выше, электростатическая энергия самого электрона не должна приниматься во внимание, так как она представляет собой самодействие. То же самое нужно сказать и об электростатической энергии протона. Энергия (13) имеет минимум при $R = R_e = \hbar^2/2me^2$, как хорошо известно [1]. Правильное минимальное значение энергии основного состояния дается (7). Соответствующее значение электростатической энергии $U_0 = -(e^2/2R_e) = -(me^4/\hbar^2)$. Это есть правильное значение. Действительно, так как $E_{0h} = U_0 + \langle T \rangle_0$, $U_0 = E_{0h} - \langle T \rangle_0$, $\langle T \rangle_0 = me^4/2\hbar^2$ (как упоминалось выше), и $E_{0h} = -(me^4/2\hbar^2)$ (смотри (7)), то $U_0 = -(me^4/\hbar^2)$.

Энергия основного состояния атома водорода вычислена здесь специальным образом, явно учитывая все слагаемые кинетической и электростатической энергий. Это было сделано для того, чтобы показать явно, что энергия основного состояния атома водорода, вычисленная здесь без включения электростатических энергий электрона и протона (то есть без учета самодействия), имеет правильное, всем известное и измеренное экспериментально значение [1].

Задача об отрицательном ионе водорода рассмотрена в [3].

## 4. Позитроний

Теперь предствим себе позитроний (атом типа водорода, состоящий из позитрона и электрона) в основном состоянии. Массы обеих частиц равны. Их заряды равны по величине и противоположны по знаку. Поэтому волновые функции обеих частиц идентичны. Получается, что в позитронии нет локального заряда и нет электростатического поля, но он обладает отрицательной электростатической энергией.

## 5. Модель однородно заряженного шара

Рассмотрим $N$ электронов, электрический заряд каждого из которых распределен равномерно в сферическом шаре радиуса $R$. Электростатическая энергия такого шара в классической электростатике есть [4] $U = 0.6e^2N^2/R$. Собственная электростатическая энергия каждого электрона равна $0.6e^2/R$. Собственная электростатическая энергия $N$ электронов это $0.6e^2N/R$. В квантовой механике эта энергия самодействия не учитывается. Поэтому в квантовой механике электростатическая энергия такого шара есть [4]:

$$U = 0.6e^2N(N-1)/R. \qquad (14)$$

В классической электростатике самодействие учитывается. Поэтому в классической электростатике вместо $N(N-1)$ в (14) стоит $N^2$.

Электростатическая энергия шара радиуса $R$, содержащего $N$ однородно распределенных в его объеме протонов выражается тем же уравнением (14).

Теперь рассмотрим шар радиуса $R$, содержащий $N$ электронов и $N$ протонов, каждый из которых распределен однородно в объеме шара. Полный заряд в каждой точке равен нулю. Электрического поля нет. Полная электростатическая энергия такого шара состоит из электростатической энергии электронов (14), электростатической энергии протонов (14) и энергии взаимодействия электронов и протонов $U_i = -1.2e^2N^2/R$. Так что полная электростатическая энергия описывается следующим быражением:

$$U_t = U_i = -1.2e^2N/R. \qquad (15)$$

Этот результат является частным случаем опубликованных ранее в [4] результатов.

В классической электростатике, как было упомянуто выше, вместо $N(N-1)$ в (14) стоит $N^2$. Следовательно, полная электростатическая энергия в классической электростатике равна нулю. В квантовой механике самодействие не учитывается и полная электростатическая энергия рассматриваемой электронейтральной системы отрицательна. Так получается потому что каждая частица производит электростатическое поле и вопреки локальной электронейтральности нужно вычесть из нуля электростатическую энергию каждой из всех частиц. Так мы приходим к отрицательной электростатической энергии.

В этом разделе рассмотрен случай, когда отсутствуют электрический заряд и поле, но имеется существенная отрицательная электростатическая энергия.

## 6. Обсуждение

Проанализировано понятие кинетической энергии в нерелятивистской квантовой механике. Кинетическая энергия является ненулевой положительной величиной во многих случаях, когда отсутствуют видимые движения и потоки. Оператор импульса содержит множителем мнимую единицу $i$. Так что квантово-механическое среднее импульса, **p**, является мнимой величиной в случае вещественой волновой функции, $\mathbf{p} = -i\hbar\int\psi\nabla\psi dv$. В то же время разлагая вещественную волновую функцию в интеграл Фурье (3), имеем: $\mathbf{p} = \hbar\int\mathbf{k}a^*_\mathbf{k}a_\mathbf{k}d\mathbf{k}$. Это вещественная величина. Отсюда следует, что для вещественной волновой функции $\mathbf{p} = 0$, так как это единственная возможность быть одновременно мнимой и вещественной величиной.

Существование положительной кинетической энергии в состояниях без потока объясняется с помощью разложения волновой функции в ряд по виртуальным плоским волнам. Это разложение приводит к представлению кинетической энергии суммой частичных кинетических энергий составляющих виртуальных плоских волн. Вещественная волновая функция является суперпозицией движущихся плоских волн. Их скорости взаимно уничтожаются, а кинетические энергии складываются.

Энергия основного состояния атома водорода вычислена специальным образом, рассматривая явно все слагаемые электростатической энергии электрона и протона и кинетической энергии электрона. Показано что правильные значения энергии основного состояния и радиуса волновой функции получаются только в случае исключения из расчетов электростатических энергий электрона и протона. Это показывает ещё раз что самодействие должно быть исключено из расчетов в квантовой механике.

Результатом отсутствия самодействия в квантовой механике являются удивительные случаи отсутствия локальных зарядов в системах, обладающих отрицательной электростатической энергией. Это происходит, в частности, в позитронии и в модели шара, содержащего равное число положительно и отрицательно заряженных элементарных частиц. Эта модель также рассмотрена в настоящей статье.

## 7. Выводы

Простой методический подход разложения волновой функции в ряд по виртуальным плоским волнам оказался полезным для понимания природы кинетической энергии частицы, описываемой вещественной волновой функцией.

Показано также, что энергия электрического поля не является во всех случаях обязательно положительной величиной.

**Благодарности**

**Литература**

[1] Л. Д. Ландау и Е. М. Лифшиц, *Квантовая Механика*, Физматгиз, Москва, 1963.
[2] Y. Kornyushin, *Facta Universitatis: Series PCT*, **2** (253) 2003.
[3] M. Ya. Amusia and Y. Kornyushin, *Eur. J. Phys.*, **21** (369) 2000.
[4] M. Ya. Amusia and Y. Kornyushin, *Contemporary Physics*, **41** (219) 2000.